\documentclass[pdftex,twocolumn,epjc3]{svjour3}  
\smartqed  % flush right qed marks, e.g. at end of proof
\RequirePackage{graphicx}
\RequirePackage{amssymb}
\RequirePackage{amsmath}
\RequirePackage{bm}
\RequirePackage{mathptmx}      
\RequirePackage{flushend}
\RequirePackage{cuted}
\RequirePackage[numbers,sort&compress]{natbib}
\RequirePackage[colorlinks,citecolor=blue,urlcolor=blue,linkcolor=blue]{hyperref}
\newcommand{\be}[1]{\begin{equation}\label{#1}}
\newcommand{\ee}{\end{equation}}
\newcommand{\ba}[1]{\begin{eqnarray}\label{#1}}
\newcommand{\ea}{\end{eqnarray}}
\newcommand{\rf}[1]{(\ref{#1})}
\newcommand{\nn}{\nonumber}

\journalname{Eur. Phys. J. C}

\begin{document}\sloppy

\title{Cosmological Perturbations Engendered by Discrete Relativistic Species%\thanksref{t1}
}


\author{Maksym Brilenkov\thanksref{e1,addr1}
        \and
 				Ezgi Canay\thanksref{e2,addr2,addr3,addr4}
        \and
       Maxim Eingorn\thanksref{e3,addr5}
        }
\thankstext{e1}{e-mail: maksym.brilenkov@astro.uio.no}
\thankstext{e2}{e-mail: e.yilmaz@hzdr.de}
\thankstext{e3}{e-mail: maxim.eingorn@gmail.com}

\institute{          Institute of Theoretical Astrophysics, University of Oslo, P.O. Box 1029 Blindern, N-0315 Oslo, Norway \label{addr1}
           \and
           Department of Physics, Istanbul Technical University, 34469 Maslak, Istanbul, Turkey \label{addr2}
           \and
          Center for Advanced Systems Understanding (CASUS), D-02826 G\"{o}rlitz, Germany \label{addr3}
           \and
          Helmholtz-Zentrum Dresden-Rossendorf, D-01328 Dresden, Germany \label{addr4}
           \and
            Department of Mathematics and Physics, North Carolina Central University,
	1801 Fayetteville St., Durham, North Carolina 27707, U.S.A. \label{addr5}                   
}

\date{Received: date / Accepted: date}

\maketitle

\begin{abstract}
Within the extension of the $\Lambda$CDM model, allowing for the presence of neutrinos or warm dark matter, we develop the analytical cosmological perturbation theory. It covers all spatial scales where the weak gravitational field regime represents a valid approximation. Discrete particles -- the sources of the inhomogeneous gravitational field -- may be relativistic. Similarly to the previously investigated case of nonrelativistic matter, the Yukawa interaction range is naturally incorporated into the first-order scalar metric corrections.
\end{abstract}

\section{Introduction}
\label{sec:1}
The extension of the conventional $\Lambda$CDM cosmological model by means of taking into account relativistic matter represents a very promising research area. Indeed, species with comparatively high peculiar velocities, such as neutrinos and warm dark matter particles, might have an important observable effect on the cosmic microwave background and large-scale structure probed by forthcoming surveys (see \cite{0603494,1111.0290,1311.5487,1710.01785,Vagnozzi1,Vagnozzi2,Batista} and references therein). The study of this effect offers an opportunity to estimate neutrino masses, predict physical properties of warm dark matter candidates and advance in solving of a number of astrophysical and cosmological challenges, such as, for instance, the well-known missing satellite problem (see, in particular, \cite{1909.05132} and references therein).

A crucial role in the investigations of the large-scale structure formation belongs to cosmological simulations. Various computer codes have been developed, for example, ``GADGET-4'' \cite{Gadget} (based on the Newtonian approximation) and ``gevolution'' \cite{gevolution} (relevant to arbitrary spatial scales and addressing the $\Lambda$CDM Universe including neutrinos). Moreover, the analytical perturbation theory covering all scales has been recently formulated in \cite{Ein1,BrilEin,Duel}, revealing the behavior of scalar and vector modes for the case of point-like nonrelativistic masses. It has been demonstrated that gravitational interactions between these masses obey the Yukawa law with the finite time-dependent screening length (for the comparison of Yukawa and Newtonian gravitational forces in a cubic simulation box see \cite{YvsN}), which emerges naturally as an analytical solution: the gravitational potential in the mentioned scheme satisfies a Helmholtz equation with the mass density perturbation as the source. The corresponding ``screening'' computer code \cite{Emrah} runs almost 40\% faster than ``gevolution'' while providing very similar outcomes in the absence of neutrinos.

In this connection, it is absolutely reasonable to generalize the analytical perturbation theory, formulated for the spatially flat Universe containing the standard $\Lambda$CDM ingredients, to the case of relativistic particles. This is our main goal here. In Section 2 we present the energy-momentum tensor components and Einstein equations for metric corrections. These equations are then solved exactly in Section 3, and two opposite limiting cases are described in Section 4. Finally, we summarize the results in concluding Section 5.

\section{Perturbed spacetime}
\label{sec:2}
We begin with presenting the perturbed Friedmann-Lema\^itre-Robertson-Walker metric in the Poisson gauge, that is 
\ba{1}
ds^2&=&a^2\left[(1+2\Phi)d\eta^2\right.\nn\\
&+&\left.2B_{\alpha}d\eta
dx^{\alpha}+\left(-\delta_{\alpha\beta}+2\Psi\delta_{\alpha\beta}\right)dx^{\alpha}dx^{\beta}\right]\, . \ea
In the above expression, $a(\eta)$ stands for the scale factor, where $\eta$ is the conformal time, and $x^\alpha$ are the comoving coordinates for $\alpha=1,2,3$. The functions $\Psi$ and $\Phi$ denote the first-order scalar perturbations whereas $B_\alpha$ is the first-order vector perturbation subject to the gauge condition $\delta^{\alpha\beta}\partial B_{\alpha}/\partial x^\beta=0$. Investigation of tensor modes is beyond the scope of our study, therefore the corresponding part in the perturbed metric is disregarded.  

Meanwhile, the components of the energy-momentum tensor for a system of point-like particles with comoving radius-vectors ${\bf r}_n$, momenta ${\bf q}_n$ and masses $m_n$ are given by (see \cite{Landau} and particularly Eqs.~(3.7), (3.10) and (3.8) in \cite{gevolution})
\ba{2}
T_0^0&=&\frac{c^2}{a^4}\sum\limits_n 
\sqrt{q^2_n + a^2m^2_n}\delta\left({\bf r}-{\bf r}_n\right)\,\nn\\ &+& 
\frac{c^2}{a^4}\overline{\sum\limits_n\frac{4q^2_n + 3a^2m^2_n}{\sqrt{q^2_n+a^2m^2_n}}\delta\left({\bf r}-{\bf r}_n\right)}\Psi\, , \ea
\be{3}
T_{\alpha}^0=-\frac{c^2}{a^4}\sum\limits_n q_n^\alpha\delta\left({\bf r}-{\bf r}_n\right)\, , \ee
\ba{4}
T_{\beta}^{\alpha}=&-&\frac{c^2}{a^4}\sum\limits_n \frac{q_n^{\alpha}q_n^{\beta}}{\sqrt{q_n^2+a^2m_n^2}}\delta\left({\bf r}-{\bf r}_n\right)\nn\\
&-&\frac{c^2}{3a^4}\overline{\sum\limits_n \frac{q^2_n\left(4q_n^2+5a^2m_n^2\right)}{\left(q_n^2+a^2m_n^2\right)^{3/2}}\delta\left({\bf r}-{\bf r}_n\right)}\delta_{\alpha\beta}\Psi\, , \ea
where $q^2_n\equiv\delta_{\alpha\beta} q_n^\alpha q_n^\beta$ and the overbar indicates spatial averaging over the comoving volume. As we allow for the contribution of relativistic species in our current approach, we do not demand that peculiar velocities of particles 
%associated with the above momenta 
are much smaller than the speed of light $c$. However, in the weak gravitational field regime, the scalar and vector perturbations are small everywhere, and thus the quantities multiplied by them are replaced by their average values, following the well-justified reasoning in \cite{Ein1,BrilEin,Chisari}. In other words, products of metric corrections with energy-momentum fluctuations are not regarded as significant sources of the inhomogeneous gravitational field. For instance, the term $\propto B_i$ in Eq.~(3.7) of \cite{gevolution} disappears in our scheme as the respective prefactor vanishes once we perform averaging. The nonzero average components
\be{5}
\overline{T_0^0}=\frac{c^2}{a^4}\overline{\sum\limits_n 
\sqrt{q^2_n + a^2m^2_n}\delta\left({\bf r}-{\bf r}_n\right)}\equiv \overline{\varepsilon}\, ,
\ee
\be{6}
\overline{T_{\beta}^{\alpha}}=
-\frac{c^2}{3a^4}\overline{\sum\limits_n \frac{q_n^2}{\sqrt{q_n^2+a^2m_n^2}}\delta\left({\bf r}-{\bf r}_n\right)}\delta_{\alpha\beta}\equiv -\overline{p}\delta_{\alpha\beta}\, 
\ee
contribute to the right-hand side of the background Friedmann equations:
\be{7}
\frac{3\mathcal{H}^2}{a^2}=\kappa \overline{\varepsilon}+\Lambda\, ,
\ee
\be{8}
\frac{2\mathcal{H}'+\mathcal{H}^2}{a^2}=-\kappa\overline{p}+\Lambda\, ,
\ee
where $\kappa\equiv 8\pi G_N/c^4$ (with $G_N$ representing the gravitational constant), $\Lambda$ is the cosmological constant, and $\mathcal{H}$ denotes the Hubble parameter defined as $\mathcal{H}\equiv a'/a\equiv (da/d\eta)/a$.

According to Eq.~\rf{1} as well as the energy-momentum tensor components presented above, Einstein equations (linearized with respect to the scalar and vector perturbations) yield   
\ba{9}
&&\triangle \Psi-3\mathcal{H}\left(\Psi'+\mathcal{H}\Phi\right)=
\,\nn\\
&&\frac{\kappa a^2}{2}\left(\frac{c^2}{a^4}\sum\limits_n 
\sqrt{q^2_n + a^2m^2_n}\delta\left({\bf r}-{\bf r}_n\right)-\overline{\varepsilon}\right)\,\nn\\
&&+ 
\frac{3\kappa a^2}{2}\left(\overline{\varepsilon}+\overline{p}\right)\Psi ,
\ea
\be{10}
\frac{1}{4}\triangle B_{\alpha} + \frac{\partial }{\partial x^{\alpha}}\left(\Psi'+\mathcal{H}\Phi\right)=-\frac{\kappa c^2}{2a^2}\sum\limits_n q_n^\alpha\delta\left({\bf r}-{\bf r}_n\right)\, ,
\ee
\ba{11}
&&\frac{\partial^2\left(\Phi-\Psi\right)}{\partial x^{\beta}\partial x^{\gamma}}-\frac{1}{3}\triangle\left(\Phi-\Psi\right)\delta_{\beta\gamma}\,\nn\\
&-&\mathcal{H}\left(\frac{\partial
	B_{\gamma}}{\partial x^{\beta}}+\frac{\partial B_{\beta}}{\partial x^{\gamma}}\right)-\frac{1}{2}\left(\frac{\partial B_{\gamma}}{\partial
	x^{\beta}}+\frac{\partial B_{\beta}}{\partial x^{\gamma}}\right)'\nn\\
&=&-\frac{\kappa c^2}{a^2}\sum\limits_n \frac{q_n^{\beta}q_n^{\gamma}-q_n^{2}\delta_{\beta\gamma}/3}{\sqrt{q_n^2+a^2m_n^2}}\delta\left({\bf r}-{\bf r}_n\right)\, ,
\ea
where the Laplace operator $\triangle\equiv\delta_{\alpha\beta}\partial^2/\partial x^\alpha\partial x^\beta$.
%%%%%%
%%%%%%%%
\section{Analytical expressions for metric corrections}
\label{sec:3}

As it follows from a scalar-vector decomposition of Eq.~\rf{11}, the difference $\Phi-\Psi$ satisfies the equation
\be{12}
\triangle\triangle(\Phi-\Psi)=-\frac{3\kappa c^2}{2a^2}\sum\limits_n \frac{q_n^{\beta}q_n^{\gamma}-q_n^{2}\delta_{\beta\gamma}/3}{\sqrt{q_n^2+a^2m_n^2}}\frac{\partial^2}{\partial x^{\beta}\partial x^{\gamma}}\delta\left({\bf r}-{\bf r}_n\right) ,
\ee
whence
\be{13}
\Phi=\Psi-\frac{3\kappa c^2}{16\pi a^2}\sum\limits_n \frac{q_n^{\beta}q_n^{\gamma}-q_n^{2}\delta_{\beta\gamma}/3}{\sqrt{q_n^2+a^2m_n^2}}\,\frac{\left(x^\gamma-x^\gamma_n\right)\left(x^\beta-x^\beta_n\right)}{\left|{\bf r}-{\bf r}_n\right|^3}\, .
\ee
Evidently, in the earlier paper \cite{Ein1} the potentials $\Phi$ and $\Psi$ were identical since quadratic momentum terms were neglected in the absence of relativistic species.

Similarly, from Eq.~\rf{10} we obtain
\be{14} \Psi'+{\mathcal H}\Phi=-\frac{\kappa c^2}{2a}\Xi\, , \ee
where
\be{15} \Xi=\frac{1}{4\pi a}\sum\limits_n\frac{q_n^\alpha(x^\alpha-x_n^\alpha)}{|{\bf r}-{\bf r}_n|^3}\, , \ee
which, when substituted into Eq.~\rf{9}, yields
\ba{16} \triangle \Psi-\frac{a^2}{\lambda^2}\Psi&=&
\frac{\kappa a^2}{2}\left(\frac{c^2}{a^4}\sum\limits_n 
\sqrt{q^2_n + a^2m^2_n}\delta\left({\bf r}-{\bf r}_n\right)-\overline{\varepsilon}\right)\,\nn\\
&-&
\frac{3\mathcal{H}\kappa c^2}{2a}\Xi\, .\ea
The Helmholtz equation~(2.27) in \cite{Ein1} for the gravitational potential  was derived in a similar manner, in a setup including nonrelativistic matter only. As was done in the mentioned paper, now we introduce the screening length $\lambda$ by means of the relationship
\be{17} \frac{a^2}{\lambda^2}=\frac{3\kappa a^2}{2}\left(\overline{\varepsilon}+\overline{p}\right)\, . \ee
It is important to emphasize that Eq.~\rf{17} can be rewritten with the help of the background Friedmann equations \rf{7} and \rf{8}:
\be{18} \lambda=\frac{a}{\sqrt{3\left(\mathcal{H}^2-\mathcal{H}'\right)}}\, ,\ee
in complete agreement with Eq.~(3.7) in \cite{Ein1}. Earlier, this universal presentation of the screening length $\lambda$, predicted in \cite{Ein1}, was confirmed for various cosmological models in \cite{EinBril,EKZ1,EKZ2,nonlinear} ($\Lambda$CDM plus extra perfect fluids with linear and nonlinear equations of state) and \cite{curvature} (nonzero spatial curvature). And now we confirm it also for the model which includes relativistic species.

The Helmholtz equation \rf{16} has the analytical solution (see \cite{Ein1} where resembling Eq.~(2.27) is analyzed):
\ba{19} \Psi&=&\frac{\kappa \lambda^2\overline{\varepsilon} }{2}-\frac{\kappa c^2}{8\pi a^2}\sum\limits_n\frac{\sqrt{q^2_n + a^2m^2_n}}{|{\bf r}-{\bf r}_n|}\exp\left(-\frac{a|{\bf r}-{\bf r}_n|}{\lambda}\right)\nn\\
&+&\frac{3\kappa c^2\lambda^2{\mathcal H}}{8\pi a^4}\sum\limits_n \frac{q^\alpha_n\left(x^\alpha-x^\alpha_n\right)}{|{\bf r}-{\bf r}_n|^3}\,\nn\\
&\times&\left[1-\left(1+\frac{a|{\bf r}-{\bf r}_n|}{\lambda}\right)\exp\left(-\frac{a|{\bf r}-{\bf r}_n|}{\lambda}\right)\right]\, .\ea

For the vector perturbation, we return to Eq.~\rf{10} and obtain
\be{20} \frac{1}{4}\triangle B_{\alpha} = -\frac{\kappa c^2}{2a}\left(\frac{1}{a}\sum\limits_n q_n^\alpha\delta\left({\bf r}-{\bf r}_n\right) - \frac{\partial\Xi}{\partial x^{\alpha}}\right)\, , \ee
whence
%(see the solution of resembling Eq.~(2.26) in \cite{Ein1})
%
\be{21} B_\alpha=\frac{\kappa c^2}{4\pi a^2}\sum\limits_{n}
\left[\frac{q_n^\alpha}{|{\bf r}-{\bf r}_n|}+\frac{q_n^\beta\left(x^\beta-x_n^\beta\right)}{|{\bf r}-{\bf r}_n|^3}\left(x^\alpha-x_n^\alpha\right)\right]\, . \ee

Finally, we write down the equation of motion for the $k$-th particle of the system as well as the connection between its momentum and peculiar velocity components (see Eq.~(3.5) and (3.4) in \cite{gevolution}, respectively):
\be{22} \left(q_k^\gamma\right)'=-\sqrt{q_k^2+a^2m_k^2}\frac{\partial\Phi}{\partial x^\gamma}-\frac{q_k^2}{\sqrt{q_k^2+a^2m_k^2}}\frac{\partial\Psi}{\partial x^\gamma}-q_k^\alpha\frac{\partial B_{\alpha}}{\partial x^\gamma}\, ,\ee
\ba{23} \tilde v_k^{\gamma}&=&\frac{q_k^\gamma}{\sqrt{q_k^2+a^2m_k^2}}+\frac{q_k^\gamma}{\sqrt{q_k^2+a^2m_k^2}}\left(2-\frac{q_k^2}{q_k^2+a^2m_k^2}\right)\Psi\nn\\
&+&\frac{q_k^\gamma}{\sqrt{q_k^2+a^2m_k^2}}\Phi + B_\gamma\, ,\ea
where $\tilde v_k^{\gamma}\equiv \left(x_k^\gamma\right)'$. All metric corrections and their spatial derivatives in Eqs.~\rf{22} and \rf{23} are calculated at the point ${\bf r}={\bf r}_k$, and the summation is over $n\neq k$.

\section{Asymptotic behavior}
\label{sec:4}

In the nonrelativistic limit, neglecting all those terms which are quadratic or higher-order in $q_n^\alpha$, one can reduce Eqs.~\rf{13} and \rf{19} to the following one:
\be{24} \Phi=\Psi=\frac{\kappa \lambda^2\overline{\varepsilon} }{2}+\Phi_{\mathrm{cold}}\, ,\ee
where
\ba{25} &&\Phi_{\mathrm{cold}}\ =\ -\frac{\kappa c^2}{8\pi a}\sum\limits_n\frac{m_n}{|{\bf r}-{\bf r}_n|}\exp\left(-\frac{a|{\bf r}-{\bf r}_n|}{\lambda}\right)\nn\\
&+&\frac{3\kappa c^2\lambda^2{\mathcal H}}{8\pi a^4}\sum\limits_n \frac{q^\alpha_n\left(x^\alpha-x^\alpha_n\right)}{|{\bf r}-{\bf r}_n|^3}\,\nn\\
&\times&\left[1-\left(1+\frac{a|{\bf r}-{\bf r}_n|}{\lambda}\right)\exp\left(-\frac{a|{\bf r}-{\bf r}_n|}{\lambda}\right)\right]\, . \ea
At the same time, in Eqs.~\rf{22} and \rf{23} one may additionally disregard those terms which contain products of $q_n^\alpha$ and the metric corrections (or their spatial derivatives):
\be{26} \left(q_k^\gamma\right)'=-am_k\frac{\partial\Phi}{\partial x^\gamma}\, ,\quad \tilde v_k^{\gamma}=\frac{q_k^\gamma}{am_k} + B_\gamma\, .\ee

In the opposite, ultrarelativistic limit, when $q_n\gg am_n$, Eqs.~\rf{13} and \rf{19} are modified as follows:
\be{27} \Phi=\frac{\kappa \lambda^2\overline{\varepsilon} }{2}+\Phi_{\mathrm{hot}}\, ,\quad \Psi=\frac{\kappa \lambda^2\overline{\varepsilon} }{2}+\Psi_{\mathrm{hot}}\, ,\ee
where
\ba{28}
&&\Phi_{\mathrm{hot}}=\Psi_{\mathrm{hot}}\,\nn\\
&-&\frac{3\kappa c^2}{16\pi a^2}\sum\limits_n \frac{q_n^{\beta}q_n^{\gamma}-q_n^{2}\delta_{\beta\gamma}/3}{q_n}\,\frac{\left(x^\gamma-x^\gamma_n\right)\left(x^\beta-x^\beta_n\right)}{\left|{\bf r}-{\bf r}_n\right|^3}\, ,
\ea
\ba{29} &&\Psi_{\mathrm{hot}}\ =\ -\frac{\kappa c^2}{8\pi a^2}\sum\limits_n\frac{q_n}{|{\bf r}-{\bf r}_n|}\exp\left(-\frac{a|{\bf r}-{\bf r}_n|}{\lambda}\right)\nn\\
&+&\frac{3\kappa c^2\lambda^2{\mathcal H}}{8\pi a^4}\sum\limits_n \frac{q^\alpha_n\left(x^\alpha-x^\alpha_n\right)}{|{\bf r}-{\bf r}_n|^3}\,\nn\\
&\times&\left[1-\left(1+\frac{a|{\bf r}-{\bf r}_n|}{\lambda}\right)\exp\left(-\frac{a|{\bf r}-{\bf r}_n|}{\lambda}\right)\right]\, . \ea
In this case Eqs.~\rf{22} and \rf{23} take the form
\ba{30} \left(q_k^\gamma\right)'&=&-q_k\frac{\partial\left(\Phi+\Psi\right)}{\partial x^\gamma}-q_k^\alpha\frac{\partial B_{\alpha}}{\partial x^\gamma}\, ,\nn\\
\tilde v_k^{\gamma}&=&\frac{q_k^\gamma}{q_k}+\frac{q_k^\gamma}{q_k}\left(\Phi+\Psi\right) + B_\gamma\, .\ea

Irrespectively of the comparison between $q_n$ and $am_n$, Eq.~\rf{21} preserves its form. It is also interesting to note that it is a common situation during the evolution of the Universe that there are simultaneously both types of particles, nonrelativistic and ultrarelativistic, or even three types, with the third one undergoing the transition (or, in simple words, cooling down).

\section{Conclusion}

Now we summarize our results obtained in the framework of the analyzed cosmological model ($\Lambda$CDM plus relativistic species):

\ 

$\bullet$ the exact analytical expressions \rf{13}, \rf{19} and \rf{21} have been derived for the first-order scalar and vector perturbations generated by discrete particles, and their asymptotic behavior has been studied;

\ 

$\bullet$ these expressions are valid at arbitrary scales and represent a direct generalization of their counterparts from \cite{Ein1} to the case of a model which includes neutrinos or warm dark matter (with massless particles also allowed);

\ 

$\bullet$ the gravitational interaction between discrete species is characterized by the finite time-dependent screening length, and its universal presentation from \cite{Ein1} has been corroborated.

\ 

The derived metric corrections along with the corresponding equations of motion are ready to be used in high-precision cosmological simulations and investigations of the role of neutrinos or warm dark matter in the structure formation. Taking into account the efficiency of the ``screening'' computer code \cite{Emrah} for nonrelativistic matter, we expect a similar positive outcome of the simulations based on the current analysis. Such promising simulations will be in the focus of our future work.

\begin{acknowledgements}
The work of Maksym Brilenkov was supported by funding from the European Union's Horizon 2020 research and innovation programme under grant agreement number 772253 (ERC; bits2cosmology).
\end{acknowledgements}


\end{document}